\documentclass[10pt,english,aps,prl,twocolumn,superscriptaddress]{revtex4-2}
\usepackage[T1]{fontenc}
\usepackage{graphicx}
\usepackage{bbm}
\usepackage{float} 
\usepackage{color}
\usepackage[usenames,dvipsnames]{xcolor}
\usepackage{amsmath} 
\usepackage{amstext}
\usepackage{latexsym}
\usepackage[colorlinks=true,citecolor=Blue,linkcolor=RubineRed,urlcolor=Blue]{hyperref}
  \hypersetup{
           breaklinks=true,   
           colorlinks=true,   
           pdfusetitle=true,  
        }
\usepackage{lipsum}
\usepackage{enumitem}
\usepackage{graphicx}

\usepackage{amsfonts}
\usepackage{epsfig}
\usepackage{dsfont}
\usepackage{mathrsfs}

\def\la{\lambda}

\def\la{\langle}
\def\ra{\rangle}

\newcommand{\beq}{\begin{equation}}
\newcommand{\eeq}{\end{equation}}
\newcommand{\beqa}{\begin{eqnarray}}
\newcommand{\eeqa}{\end{eqnarray}}
\newcommand{\normord}[1]{:\mathrel{#1}:}

\begin{document}

\title{Transitionless Quantum Driving of the Tomonaga-Luttinger Liquid}

\author{L\'eonce Dupays   \href{https://orcid.org/0000-0002-3450-1861}{\includegraphics[scale=0.05]{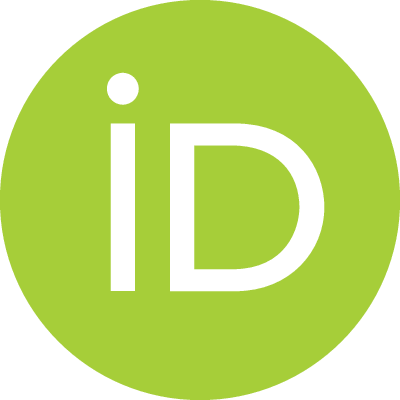}}}
\email{leonce.dupays@gmail.com}
\affiliation{Department  of  Physics  and  Materials  Science,  University  of  Luxembourg,  L-1511  Luxembourg, Luxembourg}
\author{Adolfo del Campo  \href{https://orcid.org/0000-0003-2219-2851}{\includegraphics[scale=0.05]{orcidid.eps}}}
\email{adolfo.delcampo@uni.lu}
\affiliation{Department  of  Physics  and  Materials  Science,  University  of  Luxembourg,  L-1511  Luxembourg, Luxembourg}
\affiliation{Donostia International Physics Center,  E-20018 San Sebasti\'an, Spain}

\begin{abstract}
Shortcuts to adiabaticity (STA) make the fast preparation of many-body states possible, circumventing the limitations of adiabatic strategies. 
We propose a fast STA protocol for generating interacting states in the Tomonaga-Luttinger liquid by counter-diabatic driving,  stirring the dynamics with an auxiliary control field. To this end, we exploit the equivalence between the time-dependent Tomonaga-Luttinger liquid and an ensemble of quantum oscillators with driven mass and frequency. We specify the closed-form expression of the counterdiabatic control and demonstrate its efficiency in suppressing excitations. 
\end{abstract}
\maketitle
{\it Introduction.--} Many-body quantum systems in low spatial dimensions are characterized by enhanced correlations in contrast to their three-dimensional counterparts. In one spatial dimension, the canonical Fermi liquid theory breaks down. In this setting, the Tomonaga-Luttinger liquid (TLL)  provides a universal low-energy description of interacting fermions or bosons \cite{Tomonaga1950, Luttinger1963} in terms of bosonic collective degrees of freedom  \cite{Mattis1965}. Due to its broad applicability, the TLL has been used as a test-bed for nonequilibrium quantum phenomena in theoretical and experimental studies \cite{Giamarchibook, Iucci2009, Cazalilla2011, GuanBatchelor13,Cazalilla2006,Cazalilla2016,Kinoshita2006,Rigol2007,Kaminishi2015,Langmann2017,Kaminishi2018,Yamamoto2022,Datta2023}.
Works to date predominantly focus on the nonequilibrium dynamics of the TLL  following a sudden interaction quench \cite{Dora2017,Dora2014,Dora2013,Bacsi2013,Dora2020,Bacsi2019,Pollmann2013,Kennes2013,Nessi2013,Vu2020,Wei2021}, motivated by experimental progress \cite{experimentLuttinger2008,Rueg2008}. 
In addition, understanding the nonequilibrium dynamics induced by control protocols involving finite-time driving is often necessary for the preparation of many-body states, useful in quantum technologies ranging from quantum simulation to adiabatic quantum computing \cite{Albash2018,Carcy2020,Haller_2010,Cedergren2017}.
Shortcuts to Adiabaticity (STA) offer an attractive framework for fast state preparation without the requirement for slow driving \cite{Chen10,Torrontegui13,delcampo19,GueryOdelin19}. 
A universal scheme for implementing STA relies on counter-diabatic driving (CD), which stirs the quantum evolution along the prescribed trajectory by using counterdiabatic controls \cite{Demirplak03,Demirplak05,Demirplak08,Berry09}.  
While the application of STA  by CD to many-body systems is generally challenging \cite{Campo2012a,Takahashi23}, it is largely facilitated in systems with continuous variables characterized by self-similar dynamics \cite{Chen10,delcampo11,delcampo12,Jarzynski13,delcampo13,Deffner14}. The latter naturally arises in certain systems of ultracold gases in time-dependent traps \cite{Kagan96,CastinDum96,Deng16}. In these systems,  STA schemes have been implemented with both weak and strong interactions \cite{Schaff10,Schaff11,Schaff11njp,Rohringer2015,Deng18,Deng18Sci,Diao18,Henson_2018}. 

The distribution of energy scales in the TLL hampers the engineering of finite-time driving schemes as an alternative to adiabatic protocols.
One approach relies on reverse engineering the TLL time evolution using a semiclassical sine-Gordon potential as an auxiliary control \cite{dupays2024exact}. 
As an alternative, CD provides a universal scheme for the engineering of STA. We focus on the engineering of STA in the TLL by counterdiabatic tailoring of the time-dependent interactions, paving the way to a manifold of applications ranging from optimal control  \cite{Rahmani12} to quantum thermodynamics \cite{Schmiedmayer2018, ChenWatanabe2019, Gluza21}. The TLL can describe the low-energy excitations of one-dimensional models such as the Lieb-Liniger \cite{LiebLiniger1963}, the Tonks-Girardeau gas \cite{Tonks1936, Girardeau1960}, and the Bose-Hubbard model in the continuum \cite{Cazalilla2003b}. 
 CD  in systems of continuous variables generally requires assisting the dynamics with velocity-dependent terms in the Hamiltonian description. In many important applications,  such terms can be effectively realized by a local potential, making use of a rotating frame \cite{Torrontegui13,delcampo13,Deffner14}.  However, depending on the specific system, it may not be possible to recast the CD term in a local form, i.e., as an external potential or interaction term.

This work reports STA protocols for the CD preparation of many-body states in the interaction-driven TLL. After computing the counterdiabatic controls in closed form, we establish the link between this model and an ensemble of quantum oscillators with driven mass and frequencies. We demonstrate the influence of the CD protocol on the stability of the TLL and find a range of driving speeds for which the CD protocol is relevant. We then demonstrate its efficiency in suppressing nonadiabatic energy excitations and show that the form of the CD potential varies with the nature of the interactions. \\

{\it Representations of the TLL Hamiltonian.} Let us consider the driven TLL Hamiltonian, the derivation of which is reviewed in  \cite{SM,Iucci2009,dupays2024exact}, and detailed in the supplemental material (SM) \cite{SM}, which refers to \cite{Luttinger1963,Mattis1965,Haldane1981,Mattis1974,Cazalilla2004,Giamarchibook,Lohe2008,Yang2017,Gluza_2022,Tajik2023}
\begin{align}
H_{\rm TL}(t)&=\sum_{p\neq 0}\hbar\omega(p,t)K_{0}(p)+\sum_{p\neq 0}\frac{\hbar g(p,t)}{2}\left[K_{-}(p)+K_{+}(p)\right]\nonumber\\
&-\sum_{p\neq 0}\frac{\hbar \omega(p,t)}{2},\label{eq:LuttingerHamiltonian}
\end{align}
where $v_{F}$ is the Fermi velocity. In terms of the  interaction potentials $g_{2/4}(p,t)$, $\omega(p,t)=|p|\left[v_{F}+\frac{g_{4}(p,t)}{2\pi}\right]$ and $g(p,t)=|p|\frac{g_{2}(p,t)}{2\pi}$.
%
 %
The Hamiltonian (\ref{eq:LuttingerHamiltonian}) is spanned by the generators of the $\mathfrak{su}(1,1)$ group 
\begin{align}
K_{0}(p)&=\frac{1}{2}\left[b^{\dagger}(p)b(p)+b^{\dagger}(-p)b(-p)+1\right],\\
K_{+}(p)&=b^{\dagger}(p)b^{\dagger}(-p), \quad K_{-}(p)=b(p)b(-p), \label{KpmEq}
\end{align}
that verify the commutation rules $[K_{0}(p),K_{\pm}(p)]=\pm K_{\pm}(p)$, $[K_{-}(p),K_{+}(p)]=2K_{0}(p)$ and where $b(p)$ is the annihilation operator of the noninteracting Hamiltonian ground state $|\Omega\rangle$. Alternatively, the TLL Hamiltonian can be written as a function of a phase operator $\phi(x)$ and its conjugated density fluctuation operator $\Pi(x)=\frac{\partial_{x}\theta(x)}{\pi}$, that verify the commutation relation $[\phi(x),\Pi(x')]=i\delta(x-x')$. Consider a momentum-independent interaction potential, satisfying $g_{2/4}(p,t)=\tilde{g}_{2/4}(t)$, where the tilde is used to clearly distinguish the coupling coefficients from the general $g_{2/4}(p,t)$. This corresponds to a contact interaction in real space for which the TLL Hamiltonian reads
\begin{align}
H_{\rm TL}(t)=\frac{\hbar v_s(t)}{2}\int_{0}^{L}\left[\frac{\pi}{K(t)}\Pi(x)^{2}+\frac{K(t)}{\pi}(\partial_{x}\phi(x))^{2}\right]dx,\nonumber\\
\end{align}
where the sound velocity $v_{s}$ and the Luttinger parameter $K$ are functions of the interaction potentials 
\begin{align}
\resizebox{.98\hsize}{!}{$
K(t)=\sqrt{\frac{2\pi v_{F}+\tilde{g}_{4}(t)-\tilde{g}_{2}(t)}{2 \pi v_{F}+\tilde{g}_{4}(t)+\tilde{g}_{2}(t)}},\quad v_{s}(t)=\sqrt{\left[ v_{F}+\frac{\tilde{g}_{4}(t)}{2\pi}\right]^{2}-\left(\frac{\tilde{g}_{2}(t)}{2\pi}\right)^{2}}.$}\label{eq:luttinger_parameters}\nonumber\\
\end{align}
{\it CD control and the driven TLL.} 
As a control approach, CD guides the nonadiabatic dynamics of a given system along a prescribed adiabatic trajectory \cite{Demirplak03,Demirplak05,Demirplak08,Berry09}. It removes the requirement of slow driving in adiabatic methods by introducing auxiliary controls. Specifically, consider the adiabatic trajectory associated with a system described by the time-dependent Hamiltonian $H_{0}(t)=\sum_n E_{n}(t)|n(t)\rangle\langle n(t)|$. The adiabatic evolution associated to the Hamiltonian $H_{0}(t)$ is given by the state $|\Psi_{n}(t)\rangle=e^{i\theta_{n}(t)}e^{i\gamma_{n}(t)}|n(t)\rangle$ for an initial state in the $n$-th eigenstate $|n(0)\rangle$, with the dynamical phase $\theta_{n}(t)=-\frac{1}{\hbar}\int_{0}^{t}dt'E_{n}(t')$ and the Berry phase $\gamma_{n}(t)=i\int_{0}^{t}dt'\langle n(t')|\partial_{t'}n(t')\rangle$. The CD makes the adiabatic evolution to $H_{0}(t)$ the exact evolution when the dynamics is generated by the  controlled Hamiltonian that reads $H(t)=H_0(t)+H_{\rm CD}(t)$, with the auxiliary CD term
\begin{eqnarray}
H_{\rm CD}=i\hbar \sum_{n}\left(|\partial_{t}n\rangle\langle n| -\langle n |\partial_{t} n\rangle |n\rangle\langle n| \right),
\label{HCDeq}
\end{eqnarray}
where $|n\ra=|n(t)\ra$, for short.  The instantaneous eigenstates of $H_0(t)$ can be written in terms of the evolution operator for parallel transport, $U_t=\sum_n|n(t)\ra\la n(0)|$ as $|n\rangle=U_{t}|n(0)\rangle$. As a result, the first CD term in Eq.~(\ref{HCDeq}) can  be expressed as
$
i \hbar\sum_{n}|\partial_{t}n\rangle\langle n|
=i\hbar \partial_{t}U_{t}U^{\dagger}_{t}$.
Furthermore, the Berry phase can be written as the expectation value $-i\hbar\langle n(0)|U^{\dagger}_{t}\partial_{t}U_{t}|n(0)\rangle$.

Let us focus on the case of the driven TLL when $H_{0}(t)$ is given by $H_{\rm TL}(t)$ in Eq.~(\ref{eq:LuttingerHamiltonian}). One can  diagonalize the instantaneous TLL Hamiltonian (\ref{eq:LuttingerHamiltonian}) by  applying the Bogoliubov transformation $U(\eta_{t})=\exp\left\{\sum_{p\neq 0}\frac{\eta_{t}}{2}[K_{+}(p)-K_{-}(p)]\right\}$ satisfying the condition $\tanh[2\eta_{t}]=-\frac{g(p,t)}{\omega(p,t)}$. We  define the time-dependent bosonic operators $a(p,t)=U(\eta_{t})b(p)U^{\dagger}(\eta_{t})$, satisfying
\begin{align}
a(p,t)&=\cosh(\eta_{t})b(p)-\sinh(\eta_{t})b^{\dagger}(-p),\\
b(p)&=\cosh(\eta_{t})a(p,t)+\sinh(\eta_{t})a^{\dagger}(-p,t).
\end{align}
After the Bogoliubov transformation,  the  Hamiltonian (\ref{eq:LuttingerHamiltonian})  can be brought to the diagonal form 
\begin{eqnarray}
\label{eq:diagonal_Tomonaga_Luttinger}
H_{\rm TL}(t)&=&\sum_{p\neq0}\epsilon(p,t)a^{\dagger}(p,t)a(p,t)+E_{0}(t),
\end{eqnarray}
with the ground state energy $E_{0}(t)=\frac{1}{2}\sum_{p\neq0}\left[\epsilon(p,t)-\hbar \omega(p,t)\right]$, and $\epsilon(p,t)=\hbar\sqrt{[\omega(p,t)]^{2}-[g(p,t)]^{2}}$. In other words, the instantaneous evolution of the TLL can be seen as the Bogoliubov transformation of the free TLL Hamiltonian. As noticed in \cite{Haldane1981}, the ground state energy does not diverge if $\epsilon(p,t)$ tends to $\omega(p,t)$ for high momenta. 

%
{\it Counterdiabatic control using invariant of motions.} The CD guides the nonadiabatic dynamics of the TLL along the prescribed adiabatic trajectory. The latter can be built from a unitary evolution. The supplementary gauge potential due to the nonadiabatic dynamics gives rise to the CD Hamiltonian. To formalize this idea, we exploit the theory of invariants of motions developed by Lewis and Riesenfeld for the $SU(1,1)$ dynamical group \cite{Lewis1969}, as further described in \cite{SM}. Making use of the initial invariant of motion $I^{0}=\sum_{p \neq 0}\hbar\omega_{0p}K_{0}(p)$, upon rescaling  
$I^{0}/\sigma^{2}_{t}$ with a scalar function $\sigma_{t}$, 
we build the dynamics for the interaction quench using a unitary time-evolution operator corresponding to a Bogoliubov transformation $
Q(\gamma_{p})=\exp\left\{\sum_{p\neq 0}\frac{\ln(\gamma_{p})}{2}[K_{+}(p)-K_{-}(p)]\right\}$.
%
In terms of it, the Hamiltonian reads
\begin{align}
\label{eq:main_equation}
\mathcal{H}(t)&=\sum_{p\neq0}Q(\gamma_{p})\frac{K_{0}(p)\hbar\omega_{0p}}{\sigma^{2}_{t}}Q^{\dagger}(\gamma_{p})+i\hbar\frac{\partial Q}{\partial t}Q^{\dagger},\nonumber\\
&=\hbar\sum_{p\neq 0} \bigg\{\omega_{\rm CD}(p,t)K_{0}(p)+\frac{ g_{\rm CD}(p,t)}{2}[K_{+}(p)+K_{-}(p)]\nonumber\\
&+i\frac{\dot{\gamma_{p}}}{2\gamma_{p}}\left[K_{+}(p)-K_{-}(p)\right]\bigg\}. 
\end{align}
To match the TLL, the frequency and interaction coupling 
obey
\beqa
\omega_{\rm CD}(p,t)&=&\frac{\omega_{0p}}{2}\left[\left(\frac{\gamma_{p}}{\sigma_{t}}\right)^{2}+\frac{1}{(\sigma_{t}\gamma_{p})^{2}}\right],
\nonumber\\
\\
 g_{\rm CD}(p,t)&=&\frac{\omega_{0p}}{2}\left[\frac{1}{(\sigma_{t}\gamma_{p})^{2}}-\left(\frac{\gamma_{p}}{\sigma_{t}}\right)^{2}\right].
 \eeqa
As a result, $\omega_{\rm CD}(p,t)=g_{\rm CD}(p,t)+\omega_{0p}\left(\frac{\gamma_{p}}{\sigma_{t}}\right)^{2}$. The first part of Hamiltonian \eqref{eq:main_equation} can be identified with the TLL (\ref{eq:LuttingerHamiltonian}) provided that
\beqa
K_{p}(t)=\gamma_{p}^{2},\quad  \sigma^{2}_{t}=\frac{v_{F}}{v_{s,p}(t)}, \quad\omega_{0p}=v_{F}|p|, 
\eeqa
using the momentum-dependent counterpart of (\ref{eq:luttinger_parameters})
\begin{align}
\resizebox{.98\hsize}{!}{$
K_{p}(t)=\sqrt{\frac{2\pi v_{F}+g_{4}(p,t)-g_{2}(p,t)}{2 \pi v_{F}+g_{4}(p,t)+g_{2}(p,t)}},\quad v_{s,p}(t)=\sqrt{\left[ v_{F}+\frac{g_{4}(p,t)}{2\pi}\right]^{2}-\left(\frac{g_{2}(p,t)}{2\pi}\right)^{2}}$}.\nonumber\\
\end{align}
The supplementary term  corresponds to the CD Hamiltonian
\begin{align}
H_{\rm CD}(t)=i\hbar \sum_{p\neq 0}\frac{\dot{\gamma}_{p}}{2\gamma_{p}}[K_{+}(p)-K_{-}(p)].
\end{align}
The quantum state evolution is found by considering the unitary transformation applied to the system and the dynamical phase. Consider the initial excited state for the free Hamiltonian $|\{n_{p}\},0\rangle=\bigotimes_{p\neq 0}|n_{p},0\rangle=\bigotimes_{p\neq 0}\frac{\left[b^{\dagger}(p,0)\right]^{n_{p}}}{\sqrt{n_{p}!}}|\Omega\rangle$ where $n_{p}$ is the occupation number of the $p$ mode, and $|\Omega\rangle$ the ground state of the noninteracting TLL. For a generic interacting state, $\gamma_{p}(0)\neq1$, and the evolution of the eigenstates of the TLL Hamiltonian under CD  takes the form
$$
|\Psi(t)\rangle =\exp\left[-\frac{i}{\hbar}\sum_{p\neq 0}\int_{0}^{t}\frac{\epsilon(p,0)n_{p}(0)}{\sigma^{2}_{s}}ds\right]Q(\gamma_{p})|\{n_{p}\},0\rangle.\nonumber
$$
%
Note that we can, in particular, take the ground state of the noninteracting Hamiltonian as initial state $|\{n_{p}\},0\rangle=|\Omega\rangle$; as detailed in \cite{SM}, which refers to \cite{Truax1988}.

{\it Counterdiabatic control as a time-dependent mass and frequency oscillator.} A natural physical interpretation of the CD Hamiltonian emerges by mapping the TLL model to an ensemble of quantum harmonic oscillators with time-dependent mass and frequency (\ref{eq:LuttingerHamiltonian}). To this end, we define the operators $X_{p}=\sqrt{\frac{\hbar}{2\omega_{0p}}}[b^{\dagger}(p)+b(-p)]$ and $P_{p}=i\sqrt{\frac{\hbar \omega_{0p}}{2}}[b^{\dagger}(p)-b(-p)]$. We define an effective mass $M_{p}(t)$ and frequency $\Omega_{p}(t)$ related to the interaction parameters as $|p|[v_{F}+\frac{g_{4}(p,t)}{2\pi}+\frac{g_{2}(p,t)}{2\pi}]=\frac{M_{p}(t)\Omega^{2}(p,t)}{\omega_{0p}}$ and $|p|[v_{F}+\frac{g_{4}(p,t)}{2\pi}-\frac{g_{2}(p,t)}{2\pi}]=\frac{\omega_{0p}}{M_{p}(t)}$; see further details in \cite{SM}, which refers to \cite{Torrontegui13,Mugatransi2010}. This  yields
\begin{eqnarray}
\label{TLL_HO}
&&H_{\rm TL}(t)=\sum_{p\neq 0}\frac{P_{p}P_{-p}}{2 M_{p}(t)}+\sum_{p\neq 0}\frac{1}{2}M_{p}(t)\Omega^{2}_{p}(t)X_{p}X_{-p},\nonumber\\
\end{eqnarray}
with
\begin{eqnarray}
\Omega(p,t)=v_{s,p}(t)|p|,\quad
M_{p}(t)=\frac{\omega_{0p}}{K_{p}(t)v_{s,p}(t)|p|}.
\end{eqnarray} 
As a result, the CD Hamiltonian for the TLL is given by
\begin{align}
H_{\rm CD}(t)&=\sum_{p\neq 0}\frac{\dot{K}_{p}(t)}{4K_{p}(t)}[P_{p}X_{-p}+X_{p}P_{-p}],\\
&=-\frac{1}{4}\sum_{p\neq 0}\left(\frac{\dot{\Omega}(p,t)}{\Omega(p,t)}+\frac{\dot{M}_{p}(t)}{M_{p}(t)}\right)[P_{p}X_{-p}+X_{p}P_{-p}].\nonumber
\end{align}
For each mode, the CD Hamiltonian is reminiscent of that for the harmonic oscillator with time-dependent mass and frequency \cite{SM}.  In particular, in the case $g_{2}(p,t)=g_{4}(p,t)$, $M_{p}(t)=1$, and one recovers the CD term for the time-dependent harmonic oscillator. Note that our result applies to any system described by the $SU(1,1)$ algebra \cite{Zhang_2022,AndreiRybin_1998}. This exact expression admits an intuitive interpretation given the explicit form of the $K_{\pm}(p)$ in Eq.~(\ref{KpmEq}). Indeed, the CD term in the TLL is equivalent to the sum of the generator of a two-mode squeezing, acting on the modes annihilated by $b(p)$ and $b(-p)$, respectively. For its further characterization, we resort to bosonization. 

{\it Bosonized CD Hamiltonian.} For a momentum-independent interaction potential satisfying $g_{2}(p,t)=\tilde{g}_{2}(t)$ and $g_{4}(p,t)=\tilde{g}_{4}(t)$, the Hamiltonian \eqref{eq:main_equation} can be directly expressed as a function of the fields $\phi(x)$ and $\theta(x)$ as
\begin{eqnarray}
\mathcal{H}(t)&=&\frac{\hbar v_s(t)}{2}\int_{0}^{L}\left[\frac{\pi}{K(t)}\Pi(x)^{2}+\frac{K(t)}{\pi}(\partial_{x}\phi(x))^{2}\right]dx\nonumber\\
&-&\hbar\frac{\dot{K}(t)}{4 K(t)}\int_{0}^{L}dx\left[\Pi(x)\phi(x)+\phi(x)\Pi(x)\right],\label{eq:lutt_cd}
\end{eqnarray}
where we recognize $\{\Pi(x),\phi(x)\}$ as the generator of squeezing of the bosonized fields. The bosonized TLL Hamiltonian with the CD controls (\ref{eq:lutt_cd}) is thus a field-theoretic generalization of the generalized harmonic oscillator \cite{Funo17} with instantaneous spectrum given by
\begin{eqnarray}
\epsilon(p,t)=\hbar\sqrt{v^{2}_{s}(t)|p|^{2}-\left(\dot{K}(t)/[2K(t)]\right)^{2}}.
\end{eqnarray}
To ensure the stability of the TLL along the STA, the spectrum needs to remain real-valued \cite{SM}. For contact interactions, assuming $\tilde{g}_{2}=\tilde{g}_{4}$, this implies $\left|\dot{v}_{s}(t)/v^{2}_{s}(t)\right|<2\pi/L$, setting a lower bound to the driving speed, generalizing that in the harmonic oscillator \cite{SM,Chen10,Funo17}.

In the position representation, it is informative to represent the Hamiltonian in terms of the densities of right and left movers. The bosonic operators are linked to the densities $\rho_{s}(p)$ labeled by $s=1$ for right movers and $s=-1$ for left movers $b(p)=\sqrt{\frac{2 \pi}{L |p|}}\left[\Theta(p)\rho_{1}(-p)+\Theta(-p)\rho_{-1}(-p)\right]$ and $b^{\dagger}(p)=\sqrt{\frac{2 \pi}{L |p|}}\left[\Theta(p)\rho_{1}(p)+\Theta(-p)\rho_{-1}(p)\right]$, with the Heaviside step function $\Theta(x)=1 $ for $x>0$ and $\Theta(x)=0$ otherwise. The densities in real space are linked to the fields through $\rho_{j}(x)=\normord{\Psi^{\dagger}_{j}(x)\Psi_{j}(x)}=\frac{1}{L}\sum_{p}e^{-ipx}\rho_{j}(p)$, where $\normord{\cdots}$ is the Wick ordering. Assuming the contact interaction potential for the reference Hamiltonian satisfying $g_{2/4}(x-x',t)=\tilde{g}_{2/4}(t)\delta(x-x')$, the CD term can be implemented by tailoring the long-range interaction between the densities so that $H_{\rm CD}(t)$ equals
\begin{eqnarray}
 \frac{\hbar \pi\dot{K}(t)}{2 K(t)}\int_{0}^{L}\hspace{-0.3cm}dx\int_{0}^{L}\hspace{-0.3cm}dx'\left[\rho_{1}(x)\rho_{-1}(x')-\rho_{1}(x')\rho_{-1}(x)\right]\frac{(x'-x)}{L}.\nonumber\\
\end{eqnarray}
Further details are found in the SM, which refers to ref. \cite{zwillinger2014table}. This long-range interaction resembles the $1{\rm D}$ Coulomb interaction \cite{Dhar2017,Beau20,Yang2022}, but differs by the absence of absolute value. Note that the full CD Hamiltonian preserves the even parity symmetry.

{\it Implementing STA by CD in the driven TLL.} 
\begin{figure}[t]
\includegraphics[width=\linewidth]{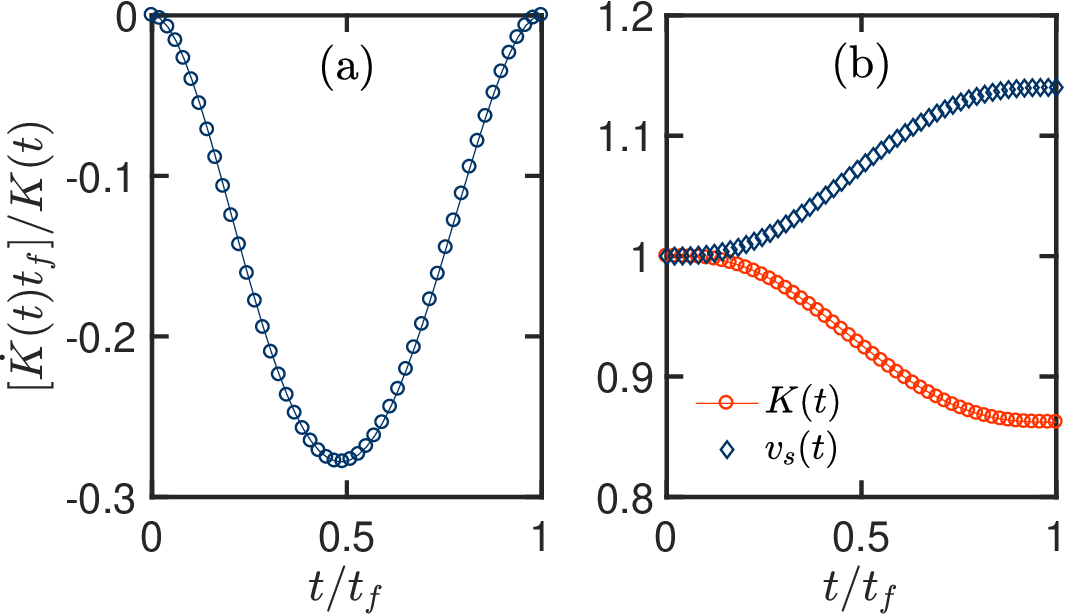}
\caption{Modulation in time of the amplitude of (a) the counter-diabatic driving Hamiltonian and (b) the Luttinger parameters for an interacting driving protocol following a fifth-order interpolating ansatz with $g_{2}(0)=g_{4}(0)=0$, $g_{2}(t_{f})=2 g_{4}(t_{f})=1v_{F}$.}\label{FigMeanEn}
\end{figure}
By way of example, let us consider the protocol for the delta interacting potential for the reference Hamiltonian. The modulation of the interaction coupling can be chosen $\tilde{g}_{2/4}(t)=\tilde{g}_{2/4}(0)+[\tilde{g}_{2/4}(t_{f})-\tilde{g}_{2/4}(0)]\mathscr{P}(t/t_{f})$ with a fifth-order polynomial ansatz $\mathscr{P}(t/t_f)=10(t/t_f)^{3}-15(t/t_f)^{4}+6(t/t_f)^{5}$. The modulation of the Luttinger parameter thus fulfills
\begin{eqnarray}
\frac{\dot{K}(t)}{K(t)}=\frac{\dot{\tilde{g}}_{4}\tilde{g}_{2}-\dot{\tilde{g}}_{2}(2\pi v_{F}+\tilde{g}_{4})}{(2\pi v_{F}+\tilde{g}_{4})^{2}-\tilde{g}^{2}_{2}}.
\end{eqnarray}
The condition on the driving speed simplifies when $\tilde{g}_{2}(t)=\tilde{g}_{4}(t)=\tilde{g}_{24}(t)$ and $\tilde{g}_{24}(0)=0$ to $t_{f}>L \frac{|\tilde{g}_{24}(t_{f}){\rm max}\dot{\mathscr{P}}(t/t_{f})|}{(2\pi v_{F})^{2}}$. The adiabatic mean energy is equal to the mean energy in the presence of CD. This result allows for preparing an interacting TLL in finite time without nonadiabatic final residual energy. Under CD, the mean energy evolves as the rescaled initial energy \cite{Funo17}
\begin{eqnarray}
\langle \mathcal{H}(t)\rangle&=&\frac{v_{s}(t)}{v_{F}}\langle \mathcal{H}(0)\rangle,
\end{eqnarray}
with $v_{s}(t)$ given in Eq.~(\ref{eq:luttinger_parameters}). To this end, it is required that the TLL Hamiltonian parameters are modulated in time as shown in Fig. \ref{FigMeanEn}, along a finite-time interaction quench associated with the fifth-order interpolating ansatz.   More generally, any long-range interaction may be assumed for $g_{2}(p,t)/(2\pi)$ and $g_{4}(p,t)/(2\pi)$. As we next discuss, an appropriate choice of the potential leads to a different form of the CD term; quenches for momentum-dependent interactions were also studied in \cite{Langmann2017, Datta2023}.

Let us consider the Lorentzian interaction potential
\beqa
V_{2/4}(x-x',t)=\frac{\lambda(t)}{\pi}\frac{R_{0}}{(x-x')^{2}+R^{2}_{0}}, 
\eeqa
that has been previously discussed in the context of the TLL  \cite{Iucci2009,Dora2011}. In the limit of vanishing interaction range $R_{0}\rightarrow 0$, the Lorentzian interaction potential reduces to the contact potential as $V_{2/4}(x-x',t)=\lambda \delta(x-x')$.
%
%
We further note that in momentum space, the corresponding interaction strength decays exponentially as $g_{2/4}(p,t)=\lambda(t)\exp(-R_{0}|p|)$. Let us consider the case $g_{2}=g_{4}=g_{24}=\lambda(t)\exp(-R_{0}|p|)$. The long-range interaction acts as a cutoff parameter where only the momenta $|p|<\frac{1}{R_{0}}$ contribute. As a consequence, one can perform a Taylor expansion of the exponential for small $R_{0}|p|$. To match with the linear approximation of the Luttinger liquid theory, one can further neglect the higher order terms in $\mathcal{O}(|R_{0}p|^{2})$ leading to the simplified CD term, as further detailed in the SM
\begin{align}
\frac{\dot{K}(t)}{K(t)}&\approx-\frac{1}{2}\dot{\lambda}(t)\left(\frac{1}{\pi v_{F}+\lambda(t)}-\frac{R_{0}\pi v_{F}|p|}{(\pi v_{F}+\lambda(t))^{2}}\right).
\end{align}
The CD term can be expressed in the position representation as
\begin{align}
&H_{CD}(t)=i\hbar\sum_{p>0}\frac{\dot{K}(t)}{2K(t)}[K_{+}(p)-K_{-}(p)]\nonumber\\
&\approx\Delta_{1}(t)\int_{0}^{L}\hspace{-0.3cm}dx\int_{0}^{L}\hspace{-0.3cm}dx'[\rho_{1}(x)\rho_{-1}(x')-\rho_{1}(x')\rho_{-1}(x)]\frac{x'-x}{L}\nonumber\\
&+\Delta_{2}(t)\int_{0}^{L}\hspace{-0.3cm}dx\int_{0}^{L}\hspace{-0.3cm}dx'[\rho_{1}(x)\rho_{-1}(x')-\rho_{1}(x')\rho_{-1}(x)]\frac{1}{x-x'}.
\end{align}
with $\Delta_{1}(t)=-\frac{\pi\hbar }{4}\frac{\dot{\lambda}(t)}{\pi v_{F}+\lambda(t)}$ and $\Delta_{2}(t)=\frac{\hbar}{4}\frac{\dot{\lambda}(t)R_{0}\pi v_{F}}{(\pi v_{F}+\lambda(t))^{2}}$. In the limit $R_{0}\to 0$, the CD Hamiltonian recovers the result for contact interaction. The second term of the CD driving with inversely linear interaction belongs to the family of Schulz-Shastry models \cite{Schulz1998}, that can be engineered with gauge fields as detailed in the SM, which refers to \cite{Atala2014,Goldman2014}.

{\it Conclusion.} We have engineered fast-driving shortcut protocols for the nonadiabatic preparation of many-body quantum states in the TLL. To this end, we have identified the auxiliary Hamiltonian controls for transitionless quantum driving of the TLL, exploiting its analogy with an ensemble of harmonic oscillators with time-dependent mass and frequency. The auxiliary CD Hamiltonian control is the sum of generators of two-mode squeezing in terms of bosonic operators and admits a natural representation upon bosonization. 
We have 
specified the CD protocol in terms of the time-dependent TLL parameters and an external potential. The auxiliary CD Hamiltonian could be engineered through a long-range interaction between right and left movers densities that varies according to the form of the $g_{2}$ and $g_{4}$ scattering. Our results open new avenues for engineering shortcuts to adiabaticity and fast quantum control protocols using effective field theories providing an alternative to adiabatic strategies. They should find broad applications in many-particle quantum thermodynamic devices, the study of thermalization, and control of ultracold gases in tight-waveguides and one-dimensional conducting systems. \\
{\it Acknowledgements.---} We are indebted to Bal\'az D\'ora and Per Moosavi for insightful comments. We would like to thank Masahito Ueda, Federico Balducci, I\~{n}igo L. Egusquiza,  Apollonas S. Matsoukas Roubeas, Armin Rahmani, and Jing Yang for the interesting discussions. This project was supported by the Luxembourg National Research Fund (FNR Grant No. 16434093). It has received funding from the QuantERA II Joint Programme with co-funding from the European Union’s Horizon 2020 research and innovation programme.

\bibliography{Luttinger}

\setcounter{secnumdepth}{2} 
\onecolumngrid 
\newpage
\setcounter{equation}{0} \setcounter{section}{0} \setcounter{subsection}{0} \renewcommand{\theequation}{S\arabic{equation}} \onecolumngrid \setcounter{enumiv}{0}
\begin{center}
\textbf{\large{}Supplemental Material}{\large\par}
\par\end{center}
\addtocontents{toc}{\protect\thispagestyle{empty}}

\title{Transitionless Quantum Driving of the Tomonaga-Luttinger Liquid}

\author{L\'eonce Dupays   \href{https://orcid.org/0000-0002-3450-1861}{\includegraphics[scale=0.05]{orcidid.eps}}}
\affiliation{Department  of  Physics  and  Materials  Science,  University  of  Luxembourg,  L-1511  Luxembourg, Luxembourg}
\author{Adolfo del Campo  \href{https://orcid.org/0000-0003-2219-2851}{\includegraphics[scale=0.05]{orcidid.eps}}}
\affiliation{Department  of  Physics  and  Materials  Science,  University  of  Luxembourg,  L-1511  Luxembourg, Luxembourg}
\affiliation{Donostia International Physics Center,  E-20018 San Sebasti\'an, Spain}

\maketitle

\onecolumngrid

\section{Details on the derivation of the Tomonaga-Luttinger model}
The free Hamiltonian of the TLL can be expressed as [$\color{Blue}3$]
\begin{eqnarray}
H_{0}&=&\sum_{p\neq 0}\hbar v_{F} |p|b^{\dagger}(p)b(p)+\frac{\pi v_{F}}{L}(N^{2}+J^{2}),
\end{eqnarray}
 with $v_{F}$ the Fermi velocity. The bosonic operators verify the usual commutation rule $[b(p),b^{\dagger}(p')]=\delta_{p,p'}$ and their link to the densities of right and left movers $\rho_{\pm 1}(p)$ is given in the main text [$\color{Blue}2,70$]. Let us now introduce interactions in the model so that the complete Hamiltonian reads $H_{\rm TL}=H_{0}+V$, where the interaction potential takes the form $V=\sum_{i,j=\pm 1}\hat{V}_{ij}$, with
{\small
\begin{align}
\hat{V}_{ij}=\frac{\pi \hbar}{L}\sum_{p>0}\{V_{ij}(p,t)\rho_{i}(-p)\rho_{j}(p)+V_{ij}(-p,t)\rho_{i}(p)\rho_{j}(-p)\}+\frac{\pi \hbar}{L}V_{ij}(0,t)\rho_{i}(0)\rho_{j}(0),
\end{align}}
and
\begin{eqnarray}
\label{eq:interaction_potential}
V_{ij}(p,p',t)&=&\frac{1}{L}\int_{0}^{L} dx \int_{0}^{L} dx'e^{ipx} V_{ij}(x-x',t) e^{ip'x'},\\
V_{ij}(p,p',t)&=&\delta_{p,-p'}V_{ij}(p,t).
\end{eqnarray}
Even parity of the interaction potential is assumed in the following, i.e.,  $V_{ij}(p,t)=V_{ij}(-p,t)$. In particular, the interaction potential for the backscattering between the right and left movers is denoted $g_{2}(p,t)/(2\pi)=V_{1,-1}(p,t)$, while the forward scattering between right movers and left movers is denoted by $g_{4}(p,t)/(2\pi)=V_{1,1}(p,t)=V_{-1,-1}(p,t)$. Combining the noninteracting and interacting parts leads to the Hamiltonian in the main text ($\color{RubineRed}1$). The zero-momentum term defined by $C_{0}(t)=\frac{\pi \hbar v_{F}}{L}(N^{2}+J^{2})+\frac{\hbar g_{2}(0,t)}{4L}\left(N^{2}-J^{2}\right)+\frac{ \hbar g_{4}(0,t)}{4L}\left(N^{2}+J^{2}\right)$ is gauged away in Eq.~($\color{RubineRed}1$). Using normal ordering, we denoted by  $N=N_{1}+N_{-1}$  the sum of left and right movers and by $J=N_{1}-N_{-1}$ the current operator. 

To establish the field representation of the TLL, consider the field operator with right and left movers $\Psi(x)\approx e^{ik_{F}x}\Psi_{1}(x)+e^{-ik_{F}x}\Psi_{-1}(x)$ [$\color{Blue}4$]. The field can be decomposed according to a phase operator $\phi(x)$ and its conjugated operator $\Pi(x)=\frac{\partial_{x}\theta(x)}{\pi}$, through the bosonization procedure [$\color{Blue}70$], yielding in the fermionic case [$\color{Blue}72$]
\begin{eqnarray}
\Psi^{\dagger}(x)\approx\sqrt{\rho_{0}-\Pi(x)}\sum_{p}e^{i(2p+1)(\pi\rho_{0}x-\theta(x))}e^{-i\phi(x)},
\end{eqnarray}
where the average density $\rho_{0}$ is related to the Fermi momentum $\pi\rho_{0}=k_{F}$. 
It is further possible to decompose the fields in the bosonic basis 
\begin{align}
&\resizebox{.5\hsize}{!}{
$\theta(x)=-N\frac{\pi x}{L}-i\sum_{k\neq 0}\sqrt{\frac{\pi |k|}{2L}}\frac{e^{-\nu |k|/2-i k x}}{k}\left[b^{\dagger}(k)+b(-k)\right],$}\\
&\resizebox{.5\hsize}{!}{$\phi(x)=J\frac{\pi x}{L}+i\sum_{k\neq 0}\sqrt{\frac{\pi |k|}{2 L }}\frac{e^{-\nu |k|/2-i k x}}{|k|}\left[b^{\dagger}(k)-b(-k)\right],$}
\end{align}
with $\nu$ a cutoff parameter that ensures convergence, in the limit $\nu \to 0$ [$\color{Blue}70$]. The relation between the field and the bosonic basis is useful for expressing the TLL in its field representation.
\section{Theory of invariants of motion}
A rigorous derivation of the CD term relies on the theory of invariants of motion. Respectively, $I$ is an invariant of motion when  $i\hbar \frac{d I(t)}{dt}=i\hbar\frac{\partial I(t)}{\partial t}-[\mathcal{H},I(t)]=0$. Making use of the spectral decomposition of the invariant of motion $I(t)=\sum_n\lambda_n|\lambda_{n},t\rangle\langle \lambda_{n},t|$ yields the solution to the time-dependent Schr\"odinger equation with Hamiltonian $\mathcal{H}$ in the form [$\color{Blue}73$] $|\Psi_{t}\rangle=\exp\left(i\int_{0}^{t} \kappa_{n}(s) ds\right)|\lambda_{n},t\rangle,$ with $\kappa_{n}(s)=\langle \lambda_{n},s |i \frac{\partial }{\partial t}-\frac{\mathcal{H}}{\hbar}|\lambda_{n},s\rangle$. One can construct a time-dependent invariant  $I=QI^{0}Q^{\dagger}$ using a time-independent operator $I^0$ and a unitary time-evolution operator $Q$. The corresponding Hamiltonian is %
\begin{eqnarray}
\label{eq:invariant_Hamiltonian}
\mathcal{H}(t)=F(I,t)+i\hbar\frac{\partial Q}{\partial t}Q^{\dagger},
\end{eqnarray}
where $F(I,t)=I/\sigma^{2}_{t}$  and $\sigma_{t}$ a scalar time-dependent function. The dynamical phase can be expressed by direct substitution of $|\Psi_{t}\rangle$ in the Schr\"odinger equation with Hamiltonian (\ref{eq:invariant_Hamiltonian}), one obtains the phase $\kappa_{n}(s)=-\frac{1}{\hbar}\langle \lambda_{n},s|F(I,t)|\lambda_{n},s\rangle$. 
\section{ Counterdiabatic controls for the quantum oscillator with driven mass and frequency}
As the TLL is analogous to an ensemble of quantum harmonic oscillators with time-dependent mass and frequency, it is useful to study the counter-diabatic driving in the latter case. We make use of Lohe's solution of   the time-dependent Schr\"odinger equation $i\hbar\frac{d|\Psi(t)\rangle}{dt}=H|\Psi(t)\rangle$ 
with Hamiltonian [$\color{Blue}73$]
\begin{eqnarray}
\label{eq:mass_time_dependent_Hamiltonian}
H(t)=\frac{\gamma^{2}}{2m_{0}\sigma^{2}_{t}}p^{2}+\frac{1}{2}m_{0}\frac{\omega^{2}_{0}}{\sigma^{2}_{t}\gamma^{2}}x^{2}+\frac{\dot{\gamma}}{2\gamma}\{x,p\}.
\end{eqnarray}
Consider  the $n$-th eigenstate $\phi_{n}$ of the (time-independent) initial harmonic oscillator $\hat{H}=\frac{p^{2}}{2m_{0}}+\frac{1}{2}m_{0}\omega^{2}_{0}x^{2}$ with associated eigenvalue $E_{n}(0)$.
Its evolution, governed by (\ref{eq:mass_time_dependent_Hamiltonian}),  reads
\begin{eqnarray}
|\Psi_{n}(t)\rangle=\gamma^{-1/2}\exp\left(-\frac{i}{\hbar}\int_{t_{0}}^{t}\frac{E_{n}(0)ds}{\sigma^{2}_{s}}\right)\phi_{n}\left(\frac{x}{\gamma},0\right).
\end{eqnarray}
 We identify the time-dependent mass $m(t)=\frac{m_{0}\sigma^{2}_{t}}{\gamma^{2}}$ and frequency $\omega^{2}(t)=\frac{\omega^{2}_{0}}{\sigma^{4}_{t}}$, so that $\gamma=\sqrt{\frac{m_{0}\omega_{0}}{m(t)\omega(t)}}$ and $\sigma_{t}=\sqrt{\frac{\omega_{0}}{\omega(t)}}$. The Hamiltonian (\ref{eq:mass_time_dependent_Hamiltonian}) takes then the form
\begin{eqnarray}
H(t)=\frac{1}{2m(t)}p^{2}+\frac{1}{2}m(t)\omega^{2}(t)x^{2}-\frac{1}{4}\left(\frac{\dot{m}(t)}{m(t)}+\frac{\dot{\omega}(t)}{\omega(t)}\right)\{x,p\}.
\end{eqnarray}
Thus, one can identify the reference Hamiltonian to be $H_{0}(t)=\frac{1}{2m(t)}p^{2}+\frac{1}{2}m(t)\omega^{2}(t)x^{2}$ and the CD term as $H_{\rm CD}(t)=-\frac{1}{4}\left(\frac{\dot{m}(t)}{m(t)}+\frac{\dot{\omega}(t)}{\omega(t)}\right)\{x,p\}$,  generalizing the well-known result for with time-dependent frequency and constant mass [$\color{Blue}79$]. The latter is recovered for $\sigma_{t}=\gamma$ [$\color{Blue}35$]. The quantum harmonic oscillator algebra can be expressed as a function of the $SU(1,1)$ generators so that $\hat{x}=\sqrt{\frac{\hbar}{2m\omega}}(a^{\dagger}+a)$ and $\hat{p}=i\sqrt{\frac{\hbar m\omega}{2}}(a^{\dagger}-a)$, with $a$ and $a^{\dagger}$ being the canonical annihilation and creation operators, verifying the commutation rule $[a,a^{\dagger}]=1$. We define $J_{+}=\frac{1}{2}a^{\dagger 2}$, $J_{-}=\frac{1}{2}a^{2}$, and $J_{0}=\frac{1}{4}(a^{\dagger}a+aa^{\dagger})$, that verify the commutation rules $[J_{0},J_{\pm}]=\pm J_{\pm}$ and $[J_{-},J_{+}]=2J_{0}$. In terms of them, $\hat{x}^{2}=\frac{\hbar}{m_{0}\omega_{0}}(2J_{0}+J_{+}+J_{-})$, $\hat{p}^{2}=\hbar m_{0} \omega_{0}(2J_{0}-J_{+}-J_{-})$ and $\{x,p\}=i\hbar(a^{\dagger 2}-a^{2})=2i\hbar(J_{+}-J_{-})$.

By analogy, one can introduce the momentum-dependent harmonic oscillator. One can introduce the momentum-dependent pseudo-position and momentum operators
\begin{align}
X_{p}&=\sqrt{\frac{\hbar}{2\omega_{0p}}}[b^{\dagger}(p)+b(-p)],\\
X_{-p}&=\sqrt{\frac{\hbar}{2\omega_{0p}}}[b^{\dagger}(-p)+b(p)].
\end{align}
They are not Hermitian, as $X^{\dagger}_{p}=X_{-p}$. However, their product is, as $(X_{p}X_{-p})^{\dagger}=X^{\dagger}_{-p}X^{\dagger}_{p}=X_{p}X_{-p}$ and
\begin{align}
X_{p}X_{-p}=\frac{\hbar}{2\omega_{0p}}[2K_{0}(p)+K_{+}(p)+K_{-}(p)].
\end{align}
Similarly, one can introduce $P_{p}=i\sqrt{\frac{\hbar\omega_{0}}{2}}[b^{\dagger}(p)-b(-p)]$ so that
\begin{align}
P_{p}P_{-p}=\frac{\hbar \omega_{0p}}{2}[2K_{0}(p)-(K_{+}(p)+K_{-}(p))].
\end{align}
One can use these operators to express the TLL Hamiltonian as an ensemble of harmonic oscillators with time-dependent mass and frequency ($\color{RubineRed}15$).

\section{Mean Energy under counterdiabatic driving}
The nonadiabatic mean energy in the driven TLL under CD reads
\begin{eqnarray}
\langle \Psi(t)|\mathcal{H}(t)|\Psi(t)\rangle&=&\langle \Psi (0)|Q^{\dagger}\left(Q\frac{I^{0}}{\sigma^{2}_{t}}Q^{\dagger}+i\hbar\frac{\partial Q}{\partial t}Q^{\dagger}\right)Q|\Psi(0)\rangle\nonumber\\
&=&\langle \Psi(0)|\frac{I^{0}}{\sigma^{2}_{t}}|\Psi(0)\rangle+i\hbar\langle\Psi(0)|Q^{\dagger}\frac{\partial Q}{\partial t}|\Psi(0)\rangle,
\end{eqnarray}
However, the explicit evaluation shows that
\begin{eqnarray}
i\hbar Q^{\dagger}\frac{\partial Q}{\partial t}&=&i\hbar \sum_{p\neq 0}\frac{\dot{\gamma_{p}}}{2\gamma_{p}}[K_{+}(p)-K_{-}(p)].
\end{eqnarray}
As a result, 
\begin{eqnarray}
\langle \Psi(t)|\mathcal{H}(t)|\Psi (t)\rangle&=&\langle \Psi(0)|\frac{I^{0}}{\sigma^{2}_{t}}|\Psi(0)\rangle+i\hbar \sum_{p\neq 0}\frac{\dot{\gamma_{p}}}{2\gamma_{p}}\langle \Psi (0)| [K_{+}(p)-K_{-}(p)]|\Psi(0)\rangle\\
&=&\langle \Psi(0)|\frac{I^{0}}{\sigma^{2}_{t}}|\Psi(0)\rangle.
\end{eqnarray}
In short, the nonadiabatic mean energy along the CD protocol can be expressed in terms of its initial value and the scaling function $\sigma_t$.

\section{Implementation of the counter-diabatic driving term}
The CD  Hamiltonian can be conveniently written as
\begin{eqnarray}
H_{\rm CD}(t)&=&-i\hbar\sum_{p>0}\frac{\dot{K}}{2 K}[b(p)b(-p)-b^{\dagger}(p)b^{\dagger}(-p)]\\
&=&-i\hbar\sum_{p>0}\frac{\dot{K}}{2 K}\frac{2\pi}{Lp}\Big(\rho_{1}(-p)\rho_{-1}(p)-\rho_{1}(p)\rho_{-1}(-p)\Big)\\
&=&-i\hbar \sum_{p>0}\frac{\dot{K}}{2 K}\frac{2\pi}{Lp}\int_{0}^{L}dx\int_{0}^{L}dx'\Big(\rho_{1}(x)\rho_{-1}(x')-\rho_{1}(x')\rho_{-1}(x)\Big)e^{ip(x'-x)}.
\end{eqnarray}
Using the quantization of the momentum $p=\frac{2\pi}{L}n$ with $n\in \mathbbm{N}^{*}$, one can write the sum
\begin{eqnarray}
\mathcal{S}(x'-x)=\sum_{n=1}^{\infty}\frac{e^{i\frac{2\pi}{L}n(x'-x)}}{n}.
\end{eqnarray}
For the special values $x=x'$ and $|x'-x|=L$, the infinite sum diverges; $\mathcal{S}(x'-x) \to +\infty$. We would like to show that at these specific points, the function $\mathcal{S}(x'-x)$ can be approximated as a delta function. Given a test function $f(x)$, as a generalized distribution, the delta function satisfies $\int dx f(x)\delta(x)=f(0)$, along with the normalization condition $\int dx \delta(x)=1$. 
We first verify that $\mathcal{S}(r)$ is a normalizable function on its domain of definition $r\in [-L,L]$. Indeed, 
\begin{align}
\int_{-L}^{L} dr \mathcal{S}(r)=\frac{1}{i\frac{2\pi}{L}} \left(\sum_{n=1}^{\infty} \frac{e^{i 2\pi n}}{n^{2}}-\frac{e^{-i 2\pi n}}{n^{2}}\right)=\frac{L}{\pi}\sum_{n=1}^{\infty}\frac{\sin(2\pi n)}{n^{2}}.
\end{align}
Furthermore, this last series is bounded in absolute value by a convergent series
\begin{align}
\left|\sum_{n=1}^{\infty}\frac{\sin(2\pi n)}{n^{2}}\right|\leq \sum_{n=1}^{\infty}\frac{1}{n^{2}}.
\end{align}
Using the property of absolute convergence, the function $\mathcal{S}(r)$ is then normalizable on $[-L,L]$. As a consequence, we can approximate $S(r)$ to mimic a delta function at the special values $x=x'$ and $|x-x'|=L$. Note, however, that outside these points, the function is not equal to zero, so we need to add a supplementary contribution to the points $x\neq x'$. Overall, we can write $\mathcal{S}(x'-x)$ as a function of its different contributions $\mathcal{S}(0)+\mathcal{S}(L)+\mathcal{S}(-L)+\mathcal{S}(x'-x)_{x'-x\neq \{0,L,-L\}}=\delta(x'-x)+\delta(x'-x-L)+\delta(x'-x+L)+\mathcal{S}(x'-x)_{x'-x\neq \{0,L,-L\}}$. Furthermore, one can use the Euler's formula to decompose the exponential
\begin{eqnarray}
\sum_{n=1}^{\infty}\frac{e^{i\frac{2\pi}{L}n(x'-x)}}{n}=\sum_{n=1}^{\infty}\frac{\cos[\frac{2\pi}{L}n(x'-x)]}{n}+i\sum_{n=1}^{\infty}\frac{\sin[\frac{2\pi}{L}n(x'-x)]}{n}.
\end{eqnarray}
It is then useful to notice the formula [$\color{Blue}83$] for $a\in (0,2\pi)$
\begin{eqnarray}
\sum_{n=1}^{\infty}\frac{\sin(na)}{n}=\frac{\pi-a}{2}.
\end{eqnarray}
As the cosine is an even function, its contribution to the integral vanishes, and one is left with
\begin{eqnarray}
H_{\rm CD}(t)&=&-i\hbar \frac{\dot{K}}{2 K}\int_{0}^{L}dxdx'\Big[\rho_{1}(x)\rho_{-1}(x')-\rho_{1}(x')\rho_{-1}(x)\Big]\Big[\delta(x'-x)+\delta(x'-x-L)+\delta(x'-x+L)+\frac{\pi-\frac{2\pi(x'-x)}{L}}{2}\Big].\nonumber\\
\end{eqnarray}
Using again that even functions do not contribute to the integral, one finds the expression quoted in the main text. 

It is interesting to look at the effect of long-range interactions on the CD Hamiltonian form. An alternative implementation is found in using the Lorentzian potential for the long-range interaction
\begin{eqnarray}
H_{\rm CD}(t)&=&-i\hbar\Delta(t)\sum_{p>0}|p|[K_{-}(p)-K_{+}(p)],\\
&=&-i\hbar\Delta(t)\sum_{p>0}\frac{2\pi}{L}[\rho_{1}(-p)\rho_{-1}(p)-\rho_{1}(p)\rho_{-1}(-p)]\\
&=&-i\hbar\Delta(t)\frac{2\pi}{L}\int_{0}^{L}dx\int_{0}^{L}dx'[\rho_{1}(x)\rho_{-1}(x')-\rho_{1}(x')\rho_{-1}(x)]\sum_{p>0}\frac{e^{ip(x'-x)}}{2}\\
&=&-i\hbar\Delta(t)\frac{2\pi}{L}\int_{0}^{L}dx\int_{0}^{L}dx'[\rho_{1}(x)\rho_{-1}(x')-\rho_{1}(x')\rho_{-1}(x)]\sum_{p>0}\frac{\left(e^{ip(x'-x)}-e^{-ip(x'-x)}\right)}{2}\\
&=&-i\hbar\Delta(t)\frac{2\pi}{L}\int_{0}^{L}dx\int_{0}^{L}dx'[\rho_{1}(x)\rho_{-1}(x')-\rho_{1}(x')\rho_{-1}(x)]\
\sum_{n=1}^\infty\frac{\left(e^{i\frac{2\pi n}{L}(x'-x)}-e^{-i\frac{2\pi n}{L}(x'-x)}\right)}{2}\\
&=&\hbar\Delta(t)\frac{\pi}{L}\int_{0}^{L}dx\int_{0}^{L}dx'[\rho_{1}(x)\rho_{-1}(x')-\rho_{1}(x')\rho_{-1}(x)]\frac{1}{\tan[\frac{\pi}{L}(x'-x)]}\\
&\approx&-\hbar\Delta(t)\int_{0}^{L}dx\int_{0}^{L}dx'[\rho_{1}(x)\rho_{-1}(x')-\rho_{1}(x')\rho_{-1}(x)]\left[\frac{1}{(x-x')}-\frac{\pi^2}{3L^2}(x-x')+\mathcal{O}(L^{-4})\right],
\end{eqnarray}
leading to the result presented in the main text.
%
\section{Spectrum through the Bogoliubov transformation}
The Tomonaga-Luttinger liquid with $\rm CD$ Hamilontian is written as  
\begin{eqnarray}
\mathcal{H}(t)&=&\hbar\sum_{p\neq 0}\omega_{\rm CD}(p,t)K_{0}(p)+\hbar \sum_{p\neq 0}\alpha(t)K_{+}(p)+\hbar \sum_{p\neq 0}\alpha^{*}(t)K_{-}(p),\\
\alpha(t)&=&\frac{g_{\rm CD}(p,t)}{2}+i\frac{\dot{\gamma_{p}}}{2\gamma_{p}}.
\end{eqnarray}
One can set 
\begin{eqnarray}
b(p)=u d(p)+v d^{\dagger}(-p).
\end{eqnarray}
The bosonic commutation relation $[b(p),b^{\dagger}(p')]=\delta_{p,p'}$ enforces the condition
\begin{eqnarray}
|v|^{2}-|u|^{2}=1,
\end{eqnarray}
leading to the parametrization
\begin{eqnarray}
u&=&e^{i\theta_{1}}\cosh(r),\\
v&=&e^{i\theta_{2}}\sinh(r).
\end{eqnarray}
In the new basis, the Hamiltonian becomes 
\begin{eqnarray}
\mathcal{H}(t)&=&\hbar \sum_{p\neq0}\left[\frac{\omega_{\rm CD}(p,t)}{2}(|u|^{2}+|v|^{2})+\alpha u^{*}v^{*}+\alpha^{*}uv \right]d^{\dagger}(p)d(p)\\
& &+\hbar \sum_{p\neq 0}\left[\omega_{\rm CD}(p,t)uv^{*}+\alpha (v^{*})^{2}+\alpha^{*}u^{2}\right]d(-p)d(p)\\
& &+\hbar \sum_{p\neq 0}\left[\frac{\omega_{\rm CD}(p,t)}{2}(|u|^{2}+|v|^{2})+\alpha v^{*}u^{*}+\alpha^{*}uv\right]d(-p)d^{\dagger}(-p)\\
& &+\hbar \sum_{p\neq 0}\left[\omega_{\rm CD}(p,t)u^{*}v+\alpha (u^{*})^{2}+\alpha^{*}v^{2}\right]d^{\dagger}(p)d^{\dagger}(-p).\nonumber\\
\end{eqnarray}
The diagonalization condition is given by
\begin{eqnarray}
0=\omega_{\rm CD}(p,t)e^{i(\theta_{2}-\theta_{1})}\frac{1}{2}\sinh(2r)+|\alpha| e^{i\phi_{\alpha}} e^{-2i\theta_{1}}\cosh^{2}(r)+|\alpha| e^{-i\phi_{\alpha}}e^{2i\theta_{2}}\sinh^{2}(r),
\end{eqnarray}
with $\theta_{1}=\theta_{2}$ with $\theta_{1}=\phi_{\alpha}/2$ and
\begin{eqnarray}
\tanh(2r)=-\frac{2|\alpha|}{\omega_{\rm CD}(p,t)}.
\end{eqnarray}
The previous relation is obeyed only if $\left|2|\alpha|/\omega_{\rm CD}(p,t)\right|<1$. The spectrum can be written as 
\begin{eqnarray} 
\epsilon(p,t)&=&\omega_{\rm CD}(p,t){\rm sech}(2r)=\sqrt{\omega^{2}_{\rm CD}(p,t)-g^{2}_{\rm CD}(p,t)-\left(\frac{\dot{\gamma_{p}}}{\gamma_{p}}\right)^{2}}\\
&=&\sqrt{\frac{\omega^{2}_{0p}}{\sigma^{4}_{t}}-\left(\frac{\dot{\gamma_{p}}}{\gamma_{p}}\right)^{2}}\\
&=&\sqrt{v^{2}_{s}|p|^{2}-\left(\frac{\dot{K}}{2 K}\right)^{2}},
\end{eqnarray}
where we used $\cosh[{\rm arctanh}(x)]=\frac{1}{\sqrt{1-x^{2}}}$ for $|x|<1$.
\section{Stability criteria for the TLL}
For the TLL to be stable, the energy spectrum should be bounded from below 
\begin{eqnarray}
\sqrt{v^{2}_{s}(t)|p|^{2}-\left(\frac{\dot{K}(t)}{2 K(t)}\right)^{2}}>0.
\end{eqnarray}
Even though small momenta are more prone to lead to an imaginary spectrum in the presence of CD, a small enough amplitude for the CD Hamiltonian may not lead to a divergence of the dynamics. Indeed, the minimal value for the momentum is $p=\frac{2\pi}{L}$. Hence,
\begin{eqnarray}
\left|v_{s}(t)\left(\frac{2\pi}{L}\right)\right|>\left|\frac{\dot{K}(t)}{2 K(t)}\right|,
\end{eqnarray}
sets a bound on the speed at which the process can be performed for finite-size systems. In the case of the delta-interacting potential $\tilde{g}_{2}(t)=\tilde{g}_{4}(t)=\tilde{g}_{24}(t)$, the condition simplifies to 
\begin{align}
\left|\frac{v_{F}[\dot{\tilde{g}}_{24}(t)/2\pi]}{v^{3}_{s}(t)}\right|=\left|\frac{\dot{v}_{s}(t)}{v^{2}_{s}(t)}\right|<\frac{2\pi}{L}.
\end{align}
This is reminiscent of  the adiabatic condition for the time-dependent harmonic oscillator established by Lewis and Riesenfeld [$\color{Blue}77$]
\begin{align}
\left|\frac{\dot{\Omega}(p,t)}{\Omega^{2}(p,t)}\right|=\left|\frac{\dot{v}_{s}(t)}{v^{2}_{s}(t)|p|}\right|\ll 1.
\end{align}
Thus, the same control parameter governs the onset of adiabaticity under slow driving and the breaking of the TLL description in a STA by CD.
The driving regime of interest for the CD protocol spans the range of driving speeds between these two limits. We explore this regime considering the previous experimental implementation of the TLL in ultracold atoms  [$\color{Blue}74\!-\!76$]. In [$\color{Blue}74$], the authors measured  a sound velocity $v_{s}\approx 2.21\ {\rm \mu m/ms}$  and considered a system size $L=10-100\ {\rm \mu m}$. In [$\color{Blue}76$], the authors considered $L\approx 50\ {\rm \mu m}$ and ${v}_{s}\approx 2.04 \ {\rm \mu m/ms}$. By dimensional analysis $|\dot{v}_{s}/v^{2}_{s}|\approx 1/(t_{f}v_{s})$, and the quench time needs to be larger than 
$t_{f}>\frac{L}{2\pi v_{s}}
\approx 3.90 \ {\rm ms}$.
The adiabatic quench has to be very large compared to this time so that the regime of interest of the counterdiabatic protocol is between $3.9 {\rm ms}$ and $39 {\rm ms}$  for the given experimental values. The sound velocity was computed with the formula ${v}_{s}=\sqrt{g_{\rm 1D}{n}_{\rm 1D}/m}$ [$\color{Blue}76$] with the $1D$ interaction strength $g_{\rm 1D}=\hbar \omega_{\perp}a_{s}\frac{2+3 a_{s}n_{\rm 1D}}{1+2 a_{s}n_{\rm 1D}}$, the three dimensional scattering length $a_{s}=5.2\ {\rm nm}$, the mass $m=1.44\times 10^{-25}\ {\rm kg}$ of the $^{87}{\rm Rb}$ atom, the transverse trapping frequency $\omega_{\perp}=2\pi\times 1.4\ {\rm kHz}$, and the density $n_{\rm 1D}\approx 70\ {\rm \mu m} ^{-1}$. 

\section{CD term for the long-range interaction quench of Lorentz form \label{CD_Lorentz}}
We consider the modulation of the interaction strength of the form $g_{2/4}(p,t)=\lambda(t)\exp(-R_{0}|p|)$, so that the amplitude of the CD driving term reads
\begin{align}
\frac{\dot{K}(t)}{K(t)}&=-\frac{1}{2}\frac{\dot{g}_{24}}{\pi v_{F}+g_{24}}.
\end{align}
To remain in the Luttinger description, one can linearize the interaction. The momentum cutoff is equivalent to assume that only momenta $|p|\ll \frac{1}{R_{0}}$ contribute. One can Taylor expand the previous expression
\begin{align}
\frac{\dot{K}(t)}{K(t)}&\approx-\frac{1}{2}\dot{\lambda}(t)\left(\frac{1}{\pi v_{F}+\lambda(t)}-\frac{R_{0}\pi v_{F}|p|}{(\pi v_{F}+\lambda(t))^{2}}+\mathcal{O}(R^{2}_{0}|p|^{2})\right).
\end{align}
As a consequence, the CD term is divided into two parts: one contribution stemming from the short-range interaction and one linear term in momentum coming from the long-range Lorentzian interaction. Also, the momentum-independent term leads to a linear interaction between densities of right and left movers in real space, and the linear term in momentum leads to the inversely linear interaction in real space. Finally, one can write the CD Hamiltonian in real space as
\begin{eqnarray}
H_{CD}(t)&=&i\hbar\sum_{p>0}\frac{\dot{K}(t)}{2K(t)}[K_{+}(p)-K_{-}(p)]\nonumber\\
&=&-\frac{i\hbar }{4}\frac{\dot{\lambda}(t)}{\pi v_{F}+\lambda(t)}\sum_{p>0}[K_{+}(p)-K_{-}(p)]+\frac{i\hbar}{4}\frac{\dot{\lambda}(t)R_{0}\pi v_{F}}{(\pi v_{F}+\lambda(t))^{2}}\sum_{p>0}|p|[K_{+}(p)-K_{-}(p)]\\
&\approx&-\frac{\pi\hbar }{4}\frac{\dot{\lambda}(t)}{\pi v_{F}+\lambda(t)}\int_{0}^{L}dx\int_{0}^{L}dx'[\rho_{1}(x)\rho_{-1}(x')-\rho_{1}(x')\rho_{-1}(x)]\frac{x'-x}{L}\nonumber\\
& & +\frac{\hbar}{4}\frac{\dot{\lambda}(t)R_{0}\pi v_{F}}{(\pi v_{F}+\lambda(t))^{2}}\int_{0}^{L}dx\int_{0}^{L}dx'[\rho_{1}(x)\rho_{-1}(x')-\rho_{1}(x')\rho_{-1}(x)]\frac{1}{x-x'}.
\end{eqnarray}
\section{Implementation of the CD term with gauge fields \label{app:gauge_field}}
In this section, we detail the implementation of the CD Hamiltonian with gauge fields. Let us consider the gauge-field transformation of the Luttinger liquid 
\begin{align}
\mathcal{H}(t)&=\frac{\hbar v_{s}(t)}{2}\int_{0}^{L}\left[\frac{\pi}{K(t)}\tilde{\Pi}(x)^{2}+\frac{K(t)}{\pi}(\partial_{x}\phi(x))^{2}\right]dx,
\end{align}
with $\tilde{\Pi}(x)=\Pi(x)-A(x)$ the shifted momentum, and the long-range gauge field of strength $\nu$
\begin{align}
A(x)&=\nu \int dz \frac{\partial_{z}\phi(z)}{z-x}.
\end{align}
A detailed expansion of the Hamiltonian reads
\begin{eqnarray}
\mathcal{H}(t)&=&\frac{\hbar v_{s}(t)}{2}\int_{0}^{L}\left[\frac{\pi}{K(t)}\Pi(x)^{2}+\frac{K(t)}{\pi}(\partial_{x}\phi(x))^{2}\right]dx\nonumber\\
& &+\frac{\hbar v_{s}(t)}{2}\frac{\pi}{K(t)}\nu^{2}\int dz\partial_{z}\phi(z)\int dy\partial_{y}\phi(y)\int_{0}^{L}dx\frac{1}{z-x}\frac{1}{y-x}\nonumber\\
& &-\frac{\hbar v_{s}(t)}{2}\frac{\pi}{K(t)}\nu\int_{0}^{L}dx\int_{0}^{L} dz\frac{1}{z-x}\left[\Pi(x)\partial_{z}\phi(z)+\partial_{z}\phi(z)\Pi(x)\right].
\end{eqnarray}
The second term in the previous equation vanishes due to the integration over orthogonal functions. To proceed further, one can express the fields in terms of the densities of left and right movers
\begin{align}
\theta(x)&=-N\frac{\pi x}{L}-i\sum_{k\neq 0}\frac{e^{-ikx}}{k}\left[\rho_{1}(k)+\rho_{-1}(k)\right],\\
\phi(x)&=J\frac{\pi x}{L}+i\sum_{k\neq 0}\frac{e^{-ikx}}{|k|}{\rm sgn}(k)\left[\rho_{1}(k)-\rho_{-1}(k)\right].
\end{align}
Finally, in omitting the constant part, the expression of the derivative of the fields is given by
\begin{align}
\partial_{x}\theta(x)&=-\sum_{k\neq 0}e^{-ikx}\left[\rho_{1}(k)+\rho_{-1}(x)\right]=-L[\rho_{1}(x)+\rho_{-1}(x)],\\
\partial_{x}\phi(x)&=\sum_{k\neq 0}e^{-ikx}\left[\rho_{1}(k)-\rho_{-1}(k)\right]=L[\rho_{1}(x)-\rho_{-1}(x)].
\end{align}
As a consequence, the interacting term reads as follows
\begin{align}
\frac{\hbar v_{s}(t)}{2}\frac{\pi}{K(t)}\frac{L^{2}}{\pi}\int_{0}^{L}dx\int_{0}^{L}dz\frac{1}{z-x}\left[2\rho_{1}(x)\rho_{1}(z)-2\rho_{-1}(x)\rho_{-1}(z)-2\rho_{1}(x)\rho_{-1}(z)+2\rho_{1}(z)\rho_{-1}(x)\right].
\end{align}
The density-density interaction between strictly right movers or strictly left movers is symmetric under permutation of the positions $x$ and $z$, and the antisymmetric contribution of the interaction potential $1/(z-x)$ leads to a null integration. Thus, one is left with 
\begin{align}
\hbar v_{s}(t)\frac{L^{2}}{K(t)}\nu\int_{0}^{L}dx\int_{0}^{L}dz\frac{1}{z-x}[\rho_{1}(z)\rho_{-1}(x)-\rho_{1}(x)\rho_{-1}(z)].
\end{align}
This Hamiltonian contributes to the CD Hamiltonian for a Lorentzian interaction in the position space. By identification with the amplitude of the CD driving term in the manuscript, one obtains 
\begin{align}
\Delta_{2}(t)=\frac{1}{4}\frac{\dot{\lambda}(t)R_{0}\pi v_{F}}{(\pi v_{F}+\lambda(t))^{2}}=-v_{s}(t)\frac{L^{2}}{K(t)}\nu.
\end{align}
Furthermore, from the main text $v_{s}(t)/K(t)\approx 2(\pi v_{F}+\lambda)$. One can then relate the amplitude of the gauge field to the derivative of the cutoff parameter
\begin{align}
\nu(t)=-\frac{1}{8L^{2}}\frac{R_{0}\dot{\lambda}}{(\pi v_{F}+\lambda)^{3}}.
\end{align}
Gauge fields can be implemented in optical lattices [$\color{Blue}89$, \hspace{-0.1cm}$\color{Blue}90$], where the lattice spacing plays the role of the short distance cutoff.

\end{document}